# Random Bond Ising Systems

# on a General Hierarchical Lattice – Exact Inequalities

Avishay Efrat

and

Moshe Schwartz

School of Physics and Astronomy

Tel Aviv University,

Ramat Aviv, Tel Aviv 69978, Israel.

Random bond Ising systems on a general hierarchical lattice are considered. Interesting inequalities between eigenvalues of the Jacobian renormalization matrix at the pure fixed point are obtained. These lead to upper bounds on the crossover exponents $\{\phi_i\}$.

Exact inequalities have played an important role in the study of critical behavior in pure [1,2] and in random systems [3,4]. The purpose of the present letter is to obtain an upper bound on the crossover exponents $\{\phi_i\}$, near the pure fixed point in a random bond Ising system on a general hierarchical lattice, where the RG transformation is exact [5]. Furthermore, since some approximate RG schemes on real lattices share with the RG on hierarchical lattices (Migdal-Kadanoff [6,7] and others [8,9]) that particular property needed to prove this inequality, it must be obeyed for real lattices, at least approximately (within those schemes). In fact, since it is believed that the critical behavior of an Ising system on a real lattice can be mimicked by that behavior on a properly chosen hierarchical lattice [10-13], it may suggest that the result obtained in the following is general.

Consider a general hierarchical lattice described schematically in Fig. 1. The shaded area shown in (a) consists of a set of lattice points where some of the pairs are joined. In (b), a typical shaded area is represented. The full lines are bonds to be iterated in constructing the lattice while the dashed ones are not to be iterated. The bold lines represent the possibility for some of the bonds to be strengthened, multiplied by some constant. All three types of bonds carry a coupling $J_{\alpha\beta}^{12}$ (for the bond joining sites $\alpha$ and $\beta$), governed by a distribution $P(J_{\alpha\beta}^{ij})$ that is identical for all bonds (Note that <u>one</u> of the members of the pair $\alpha\beta$ may be either 1 or 2).

The renormalized coupling is given by

$$\widetilde{J}_{12} = f\{J_{\alpha\beta}^{12}\}, \tag{1}$$

where $f$ depends only on couplings associated with the pair of sites (1,2) (the shaded area, Fig 1). This implies that $\widetilde{J}_{ij}$ and $\widetilde{J}_{lm}$ are not correlated if the pairs $(i,j)$ and $(l,m)$ are not identical. The renormalized distribution, $\widetilde{P}(\widetilde{J}_{ij})$, is given by

$$\widetilde{P}(\widetilde{J}_{ij}) = \int \prod_{\alpha\beta} dJ_{\alpha\beta}^{ij} P(J_{\alpha\beta}^{ij}) \delta[\widetilde{J}_{ij} - f\{J_{\gamma\delta}^{ij}\}] \tag{2}$$

and may serve to derive an infinite set of equations for the renormalized moments.

We denote

$$\mu = \langle J_{\alpha\beta} \rangle \tag{3}$$

and

$$\Gamma_i = \langle (\delta J_{\alpha\beta})^i \rangle, \qquad i \geq 2 \tag{4}$$

(there is no need to assume that the distribution is symmetric with respect to $\delta J_{\alpha\beta} \to -\delta J_{\alpha\beta}$ but, of course, if we start with a symmetric distribution it will remain symmetric under renormalization).

The recursion equations for the moments read

$$\widetilde{\mu} = F[\mu, \Gamma_2, \Gamma_3, ...] \tag{5}$$

and

$$\widetilde{\Gamma}_i = G_i[\mu, \Gamma_2, \Gamma_3, ...] \tag{6}$$

We assume the existence of a pure ferromagnetic fixed point at $J^* > 0$. This implies that

$$J^* = F[J^*, 0, 0, ...] \tag{7}$$

and

$$\Gamma_i^* = 0. \tag{8}$$

The linearized RG transformations near the pure fixed point have the form

$$\delta\widetilde{\mu} = \frac{\partial F}{\partial \mu}(J^*, 0, 0, ...)\delta\mu + \sum_{j=2}^{\infty} \frac{\partial F}{\partial \Gamma_j}(J^*, 0, 0, ...)\Gamma_j \tag{9}$$

and

$$\widetilde{\Gamma}_i = \frac{\partial G_i}{\partial \mu}(J^*, 0, 0, ...)\delta\mu + \sum_{j=2}^{\infty} \frac{\partial G_i}{\partial \Gamma_j}(J^*, 0, 0, ...)\Gamma_j. \tag{10}$$

It is clear that $\tilde{\Gamma}_i = 0$ for all $i$ if all the $\Gamma_i$'s vanish. Therefore,

$$\frac{\partial G_i}{\partial \mu}(J^*,0,0,...) = 0. \tag{11}$$

Furthermore, when all the $\Gamma_i$'s are small, $|\Gamma_i|$ is of the order of $\Gamma_2^{i/2}$ or less (the odd moments may be identically zero, for example). Therefore, it is clear that

$$\frac{\partial G_i}{\partial \Gamma_j}(J^*,0,0,...) = 0 \quad \text{for} \quad j < i. \tag{12}$$

The Jacobian transformation matrix at the pure fixed point is thus triangular and its eigenvalues are just the diagonal elements of the matrix.

We prove next the following properties of the eigenvalues $\{\lambda_i\}$:

(a) All the eigenvalues are positive.

(b) $\lambda_{i+1} < \lambda_i$. \hfill (13)

(c) The eigenvalues obey a convexity condition

$$\lambda_i \lambda_j \geq \lambda_{i+j}. \tag{14}$$

Proof: Some straightforward algebra is needed to obtain the diagonal elements of the matrix in terms of the RG transformation of Eq. (1). We find

$$\lambda_i = \sum_{(\alpha,\beta)} \left[\frac{\partial f}{\partial J_{\alpha\beta}}(J^*,J^*,...)\right]^i, \tag{15}$$

where the sum is over all bonds $(\alpha,\beta)$ associated with the pair (1,2) and the partial derivative is taken at the point where all those couplings equal $J^*$.

Properties (a) and (c) are proven by showing that

$$\frac{\partial f}{\partial J_{\alpha\beta}}(J^*,J^*,...) \geq 0. \tag{16}$$

Consider the system of spins in Fig. 1 (without the implied iteration that produces the hierarchical lattice). The interaction among the spins can be described in terms of a Hamiltonian, $H\{\sigma_\alpha;\sigma_1,\sigma_2\}$, that depends on the spins $\{\sigma_\alpha\}$ internal to the shaded region and the external spins $\sigma_1$ and $\sigma_2$. An effective interaction between $\sigma$ and $\sigma_2$, $H_{\text{eff}}$, is given by

$$H_{\text{eff}} = -\ln \operatorname*{tr}_{\{\sigma_\alpha\}} e^{-H\{\sigma_\alpha;\sigma_1,\sigma_2\}}. \tag{17}$$

The most general form of an even Ising Hamiltonian depending on two spins is

$$H_{\text{eff}}(\sigma_1,\sigma_2) = C - K\sigma_1\sigma_2. \tag{18}$$

The coupling $K$ is nothing but the renormalized coupling $\widetilde{J}_{12}$ defined in Eq. (1). Now, in the vicinity of the point where $J^{12}_{\gamma\delta} = J^*$ for all $(\gamma,\delta)$, the Hamiltonian $H\{\sigma_\alpha;\sigma_1,\sigma_2\}$ is ferromagnetic and therefore the GKS inequalities apply [14,15]. Thus

$$\frac{\partial \langle \sigma_1\sigma_2 \rangle_H}{\partial J_{\alpha\beta}} \geq 0, \tag{19}$$

where $\langle ... \rangle_H$ denotes thermal average with respect to $H\{\sigma_\alpha;\sigma_1,\sigma_2\}$. (The derivative is taken at the point where all the $\widetilde{J}_{\gamma\delta}$'s equal $J^*$ but this is not crucial. For Eq. (19) to hold, it is enough that all the couplings are ferromagnetic.) The correlation $\langle \sigma_1\sigma_2 \rangle_H$ is given by (Eqs. (18), (1))

$$\langle \sigma_1\sigma_2 \rangle_H = \operatorname{tgh}(\widetilde{J}_{12}). \tag{20}$$

Now,

$$\frac{\partial \langle \sigma_1\sigma_2 \rangle_H}{\partial J_{\alpha\beta}} = \frac{1}{\cosh^2(\widetilde{J}_{12})}\frac{\partial \widetilde{J}_{12}}{\partial J_{\alpha\beta}}, \tag{21}$$

so that, from Eq. (19), it follows that $\partial \widetilde{J}_{12}/\partial J_{\alpha\beta} \geq 0$.

Property (b) is now shown to hold by proving that $\partial \widetilde{J}_{12}/\partial J_{\alpha\beta} < 1$ at any finite temperature. From Eq. (18) it is clear that the renormalized coupling $\widetilde{J}_{12}$ is given by

$$\widetilde{J}_{12} = \frac{1}{4} \operatorname*{tr}_{\sigma_1,\sigma_2} \sigma_1 \sigma_2 \ln \operatorname*{tr}_{\{\sigma_\alpha\}} e^{-H\{\sigma_\alpha;\sigma_1,\sigma_2\}}, \tag{22}$$

so that

$$\frac{\partial \widetilde{J}_{12}}{\partial J_{\alpha\beta}} = \frac{1}{4} \operatorname*{tr}_{\sigma_1,\sigma_2} \sigma_1 \sigma_2 \langle \sigma_\alpha \sigma_\beta \rangle_{12}, \tag{23}$$

where $\langle \sigma_\alpha \sigma_\beta \rangle_{12}$ is the average of $\sigma_\alpha \sigma_\beta$ with respect to $H\{\sigma_\alpha;\sigma_1,\sigma_2\}$ with $\sigma_1$ and $\sigma_2$ fixed. Now,

$$\frac{\partial \widetilde{J}_{12}}{\partial J_{\alpha\beta}} \leq \frac{1}{4} \operatorname*{tr}_{\sigma_1,\sigma_2} |\sigma_1 \sigma_2| \left| \langle \sigma_\alpha \sigma_\beta \rangle_{12} \right| = \frac{1}{4} \operatorname*{tr}_{\sigma_1,\sigma_2} \left| \langle \sigma_\alpha \sigma_\beta \rangle_{12} \right| \leq 1. \tag{24}$$

The equality sign can hold only at zero temperature (infinite $J$'s).

Denoting the maximal value of $\partial \widetilde{J}_{12}/\partial J_{\alpha\beta}$ by $m < 1$ we arrive at the conclusion

$$\lambda_i \leq \lambda_1 m^{i-1}, \tag{25}$$

so that we have proven that the number of relevant interactions at the fixed point is finite. (The equality sign holds only for the Diamond Hierarchical Lattice (DHL) [5,16], an only case for which all bonds are equivalent.) It is obvious that $\lambda_1 > 1$ but there is no such limitation on $\lambda_2$. Therefore, the condition for criticality of the pure fixed point is $\lambda_2 < 1$ while for $\lambda_2 > 1$ we expect a random critical point with a different set of critical exponents. The above obvious condition for the criticality of the pure fixed point should be related somehow to a Harris criterion [17] properly defined on a hierarchical lattice. Indeed, in the special case of the DHL, it follows directly from our analysis, as was shown a long time ago [18,19] that the requirement $\lambda_2 < 1$ is equivalent to $\alpha < 0$ (with the dimension being the fractal dimension of the lattice). The connection of the condition on $\lambda_2$ with the Harris criterion on a general hierarchical lattice is not clear yet. This is mainly due to the fact that for the DHL there is only one independent eigenvalue $\lambda_1$, while in any other hierarchical lattice the number of independent eigenvalues is larger (It is equal, in fact, to the number of different sets of equivalent bonds in the shaded area connecting 1 and 2). We hope to come back to the relation with the Harris criterion in the near future.

The behavior near the fixed point in the case that $\lambda_2,...,\lambda_n > 1$ is characterized by $n-1$ crossover exponents $\phi_2,...,\phi_n$ with $\phi_i = y_i/y_1$ and $y_i = \ln\lambda_i/\ln b$, where $b$ is the rescaling factor. From Eq. (25) follows an inequality for the crossover exponents

$$\phi_i < 1 + \frac{(i-1)\ln m}{\ln b} < 1. \tag{26}$$

We wish to conclude in emphasizing that the above inequalities hold not only for exact RG transformations on hierarchical lattices but also for all other renormalization schemes (such as the MK [6,7]), in which the renormalized couplings, $\widetilde{J}_{ij}$, are not correlated and we may expect the inequalities to hold also in cases where it is clear that the correlations are not important [9].

It is tempting to speculate, that the above results are general and hold for real $d$ dimensional lattices but due to the appearance of correlations and many spin interactions under renormalization, a proof, or a disproof, seems extremely difficult.

**Acknowledgement:** This work was supported in part by a grant from the Israeli Foundation for basic research.

**Figures:**

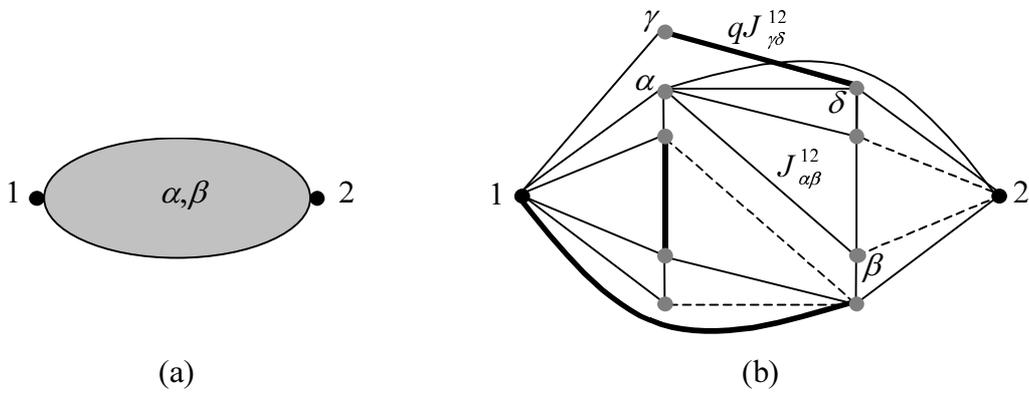

(a)  (b)

**Figure 1**

**Captions:**

1. A general hierarchical lattice is described schematically. In (a), the shaded area consists of a set of lattice points, $\alpha, \beta, ...$, where some of the pairs are joined. In (b), a typical shaded area is represented. The full lines are bonds to be iterated in constructing the lattice while the dashed ones are not to be iterated. The bold lines represent the possibility for some of the bonds to be strengthened, multiplied by some constant $q$.